\documentclass[sigconf,natbib=false,anonymous=false]{acmart}
\setcopyright{none}
\renewcommand\footnotetextcopyrightpermission[1]{}

\AtBeginDocument{%
  }

\usepackage{hyperref}
\usepackage[hang,flushmargin]{footmisc}

\usepackage{url}            
\usepackage{booktabs}       
\usepackage{amsfonts}       
\usepackage{nicefrac}       
\usepackage{microtype}      
\usepackage{amsmath}
\usepackage{enumitem}
\usepackage{graphicx}       
\usepackage{caption}
\usepackage{subcaption}
\usepackage{threeparttable}
\usepackage{color, colortbl}
\usepackage{threeparttable,booktabs}
\usepackage{multirow}
\usepackage{xspace}
\usepackage{balance}
\usepackage{soul} 
\usepackage{wasysym}
\usepackage{mdframed}

\PassOptionsToPackage{dvipsnames}{xcolor}
\usepackage{ragged2e}

\newcommand\ignore[1]{}

\RequirePackage[
  datamodel=acmdatamodel,
  style=acmnumeric,
  ]{biblatex}

\begin{document}

\title{Still Fresh? Evaluating Temporal Drift in Retrieval Benchmarks}

\author{Nathan Kuissi}
\authornote{Equal Contribution. \quad $\dagger$ Correspondence: nandan.thakur@uwaterloo.ca}
\email{ngkuissi@uwaterloo.ca}
\affiliation{%
  \institution{University of Waterloo}
  \city{Waterloo}
  \country{Canada}
}

\author{Suraj Subrahmanyan}
\authornotemark[1]
\email{s2subrah@uwaterloo.ca}
\affiliation{%
  \institution{University of Waterloo}
  \city{Waterloo}
  \country{Canada}
}

\author{Nandan Thakur$^\dagger$}
\email{nandan.thakur@uwaterloo.ca}
\affiliation{%
  \institution{University of Waterloo}
  \city{Waterloo}
  \country{Canada}
}

\author{Jimmy Lin}
\email{jimmylin@uwaterloo.ca}
\affiliation{%
  \institution{University of Waterloo}
  \city{Waterloo}
  \country{Canada}
}

\renewcommand{\shortauthors}{Kuissi et al.}
\pagestyle{empty}

\begin{abstract}
    Information retrieval (IR) benchmarks typically follow the Cranfield paradigm, relying on static and predefined corpora. However, temporal changes in technical corpora, such as API deprecations and code reorganizations, can render existing benchmarks stale. In our work, we investigate how temporal corpus drift affects FreshStack, a retrieval benchmark focused on technical domains. We examine two independent corpus snapshots of FreshStack from October 2024 and October 2025 to answer questions about LangChain. Our analysis shows that all but one query posed in 2024 remain fully supported by the 2025 corpus, as relevant documents ``migrate'' from LangChain to competitor repositories, such as LlamaIndex. Next, we compare the accuracy of retrieval models on both snapshots and observe only minor shifts in model rankings, with overall strong correlation of up to 0.978 Kendall $\tau$ at Recall@50. These results suggest that retrieval benchmarks re-judged with evolving temporal corpora can remain reliable for retrieval evaluation. We publicly release all our artifacts at \url{https://github.com/fresh-stack/driftbench}.
\end{abstract}

\ccsdesc[500]{Information systems~Evaluation of retrieval results}
\ccsdesc[300]{Information systems~Test collections}

\keywords{Temporal Analysis; Retrieval benchmarks; Evaluation}

\maketitle

\section{Introduction}
Information retrieval (IR) benchmarks rely on static test collections to evaluate and compare retrieval systems~\cite{zobel:1998, vorhees:2009}. 
Under the Cranfield paradigm, test collections are constructed once with a fixed document corpus, a set of queries, and relevance judgments. Popular IR benchmarks such as MS MARCO~\cite{nguyen:2016}, TREC Deep Learning (DL)~\cite{trec_dl:2021, craswell:2022, craswell:2023, craswell:2024}, and BEIR~\cite{thakur:2021} all follow this approach, with fixed corpus snapshots and relevance judgments created during benchmark construction.

Test collections have helped establish reusable baselines~\cite{vorhees:2000, carterette:2010, vorhees:2018}, but information changes and corpora continue to evolve and incur temporal changes~\cite{vorhees:2022, keller:2024, keller-2:2024, parry:2025}. 
Real-world document collections, such as technical documentation, are highly dynamic, as code is frequently created, deleted, and reorganized. 
Over time, the document collection snapshot captured in a benchmark diverges from the current state of the domain. 

In our work, we evaluate temporal drift in \textbf{FreshStack}, introduced by Thakur et al.~\cite{thakur2025freshstack}, using temporal snapshots of technical documentation in LangChain. 
We construct test collections using two independent corpus snapshots of ten GitHub repositories providing documentation for retrieval-augmented generation (RAG) frameworks and developer tools, by reproducing at October 2024 and extending it to October 2025. During this period, LangChain's documentation decreased by 67\% through reorganization and deprecation, while Chroma grew by 2.6$\times$ as documents migrated.

We reproduce the corpus construction and generate relevance judgments at both snapshots. 
We retrieve potentially relevant documents using a hybrid fusion by incorporating BM25~\cite{bm25}, BGE (Gemma-2)~\cite{chen:2024}, E5 Mistral (7B)~\cite{wang:2024b}, and Qwen3 (4B) Embedding~\cite{qwen_3:2025}. 
Next, we use Cohere's Command A as the judge~\cite{cohere2025command} for nugget-level support assessment~\cite{pradeep2025greatnuggetrecallautomating, 10.1145/3726302.3730316} to automatically construct test collections at both time snapshots for 203 queries in LangChain~\cite{thakur2025freshstack}. 
We explore the following three research questions in our work:

\begin{itemize}[noitemsep]
\item[{\bf RQ1}] Can existing queries be grounded in a corpus dynamically changing with time?
\item[{\bf RQ2}] How does the distribution of relevant documents across repositories change over time?
\item[{\bf RQ3}] Do model rankings remain consistent under temporal drift in technical documentation?

\end{itemize}

\noindent The contribution of this work is, to our knowledge, the first to evaluate temporal changes in a highly dynamic corpus, namely technical documentation in a niche domain. 
Prior work has evaluated temporal changes in more generic corpora, such as news~\cite{vorhees:2022}, federal documents~\cite{10.1145/1148170.1148220}, or scientific collections in LongEval~\cite{10.1007/978-3-031-88720-8_58, 10.1145/3539618.3591921}. 
Our main contribution stems from our evaluation of various retrieval systems on a dynamically changing corpus, which enables us to assess performance changes over time. 

Our experimental results demonstrate that surprisingly \emph{all but one} query remains fully grounded and answerable in both corpus snapshots, despite extensive restructuring in repositories such as LangChain. 
A major reason for this observation lies in the fact that relevant documents present in LangChain in the 2024 snapshot, are distributed across related GitHub repositories (such as LlamaIndex) in the 2025 snapshot. 
Finally, model rankings on the document retrieval task remain strongly correlated (Kendall $\tau$ = 0.978 at Recall@50), demonstrating that benchmarks can robustly evaluate retrieval systems, despite temporal changes affecting the corpus. 

\section{Background and Related Work}

Temporal drift refers to the systematic change in a corpus, queries and its associated relevance landscape over time, caused by additions, deletions, updates, or reorganization of documents.

Recent work on evaluating temporal drift in test collections, suggests that benchmarks may lose validity over time~\cite{parry:2025}, those studies primarily addresses judgment variation or query drift rather than corpus drift. 
What remains unclear is how benchmarks behave when the corpus undergoes substantial reorganization. Other works share conceptual similarity with our experiments~\cite{10.1145/3539618.3591921, keller:2024}. 
However, these explorations focus on web search data, where queries are compact and short. 
In contrast, our analysis focuses on more complex question settings, posing a strong challenge for the retrieval systems being tested. Futhermore, Cancellieri et al.~\cite{10.1007/978-3-031-88720-8_58} explore similar topics by varying both the corpus and queries across time. 
However, in this work we limit ourselves to a dynamic corpus such as technical documentation. 
In the context of coding, queries tend to be related to the same task, but due to documentation changes, implementation can vary from version to version. 
Therefore, in our context a dynamic corpus provides sufficient material for examination. 
Lastly, Soboroff~\cite{10.1145/1148170.1148220} reviews the impact of temporal changes on the dataset itself, whreas we examine retrieval system's performance, not just the benchmark.

\begin{table}[t]
\centering
\caption{LangChain GitHub repository statistics and supporting document distribution across 2024 and 2025 snapshots}
\label{tab:related_repositories}
\centering
\resizebox{0.48\textwidth}{!}{
\begin{tabular}{lrrrr}
\toprule
\multicolumn{1}{l}{\multirow{2}{*}{\textbf{GitHub Repository}}}  & \multicolumn{2}{c}{\textbf{October 2024}} & \multicolumn{2}{c}{\textbf{October 2025}} \\
\cmidrule(lr){2-3} \cmidrule(lr){4-5}
 & \textbf{\#Docs} & \textbf{Supporting (\%)} & \textbf{\#Docs} & \textbf{Supporting (\%)} \\
\midrule
\href{https://github.com/langchain-ai/langchain}{\textcolor{magenta}{langchain}} & 11,037 & 2,000 (50.9\%) & 3,628 & 921 (24.8\%) \\
\href{https://github.com/langchain-ai/langchainjs}{\textcolor{magenta}{langchainjs}} & 3,852 & 733 (18.6\%) & 3,921 & 950 (25.5\%) \\
\href{https://github.com/run-llama/llama_index}{\textcolor{magenta}{llama\_index}} & 11,627 & 634 (16.1\%) & 9,751 & 842 (22.6\%) \\
\href{https://github.com/huggingface/transformers}{\textcolor{magenta}{transformers}} & 13,580 & 173 (4.4\%) & 13,798 & 223 (6.0\%) \\
\href{https://github.com/openai/openai-cookbook}{\textcolor{magenta}{openai-cookbook}} & 537 & 160 (4.1\%) & 970 & 320 (8.6\%) \\
\href{https://github.com/chroma-core/chroma}{\textcolor{magenta}{chroma}} & 1,359 & 127 (3.2\%) & 2,949 & 245 (6.6\%) \\
\href{https://github.com/Azure-Samples/openai}{\textcolor{magenta}{azure-openai-samples}} & 413 & 49 (1.2\%) & 1,717 & 125 (3.4\%) \\
\href{https://github.com/openai/openai-python}{\textcolor{magenta}{openai-python}} & 654 & 20 (0.5\%) & 1,274 & 25 (0.7\%) \\
\href{https://github.com/Azure-Samples/azure-search-openai-demo}{\textcolor{magenta}{azure-search-openai-demo}} & 406 & 19 (0.5\%) & 1,052 & 36 (1.0\%) \\
\href{https://github.com/LangChain-ai/LangChain-nextjs-template}{\textcolor{magenta}{LangChain-nextjs-template}} & 93 & 16 (0.4\%) & 105 & 32 (0.9\%) \\
\midrule
\textbf{Total} & \textbf{43,558} & \textbf{3,931} & \textbf{39,165} & \textbf{3,719} \\
\bottomrule
\end{tabular}}
\end{table}

\section{Experimental Setup \& Details}

Evaluating retrieval systems requires judgment pools constructed from question--answer pairs. 
Furthermore, we require a sufficiently large corpus that can provide the relevant context needed to ground answers to the questions. 
Building on previous work in FreshStack~\cite{thakur2025freshstack}, we use the Stack Overflow queries originally constructed in October 2024 and extend the test collection in October 2025.\footnote{https://huggingface.co/datasets/freshstack/queries-oct-2024} 

Additionally, we focus on the \textbf{LangChain} domain for our exploration. 
While observing multiple domains may allow for a comprehensive analysis on the effects of temporal changes on retrieval performance, focusing on a single domain allows us to study temporal changes more effectively.
LangChain is a representative domain for temporal exploration, as its GitHub repository is one of the most active and rapidly changing repositories in NLP~\cite{liu2026large}. 

Our pipeline follows FreshStack~\cite{thakur2025freshstack} and automatically creates a retrieval benchmark at any time step based on the steps listed below. Each step is described in the following subsections.

\begin{enumerate}[noitemsep]
    \item  \textbf{Corpus Preparation}: We collect and chunk two versions of LangChain--related repositories for corpus preparation.
    \item \textbf{Nugget Generation}: We generate nuggets, or key facts using question and answers from Stack Overflow.
    \item \textbf{Oracle Retrieval}: We retrieve a subset of document chunks from the corpus, potentially relevant in answering the query.
    \item \textbf{Nugget-level Assessment}: Using the subset of document chunks, for every query, we determine whether each chunk supports a nugget (a key fact needed to answer the question). 
\end{enumerate}

\subsection{Corpus Preparation}

The goal here is to collect and construct a corpus using two distinct versions of LangChain related repositories, at different time snapshots.  
Moreover, we combine multiple repositories to form our corpus because a query tagged as LangChain on Stack Overflow might require context from other repositories. 
For example, a LangChain question could concern the ability to properly load a HuggingFace transformer model, or use ChromaDB. 
We provide the ten relevant repositories used to form our corpus in \autoref{tab:related_repositories}. 

To automate the procedure,  we use the main branch snapshot for all GitHub repositories as of October 2024 and October 2025. 
After providing the repository name, we search for GitHub commits prior to those dates and use the latest commit hash. 
Then, we \textbf{chunk} all documentation, code files, Jupyter notebooks, etc., with a maximum of 2048 tokens. 
To create a unique document identifier, we use the repository name, the relative file path within the local repository and the chunk start and end byte information, for disambiguation.

\begin{figure}[t!]
    \centering
    \includegraphics[trim={10 10 10 10}, clip, width=0.48\textwidth]{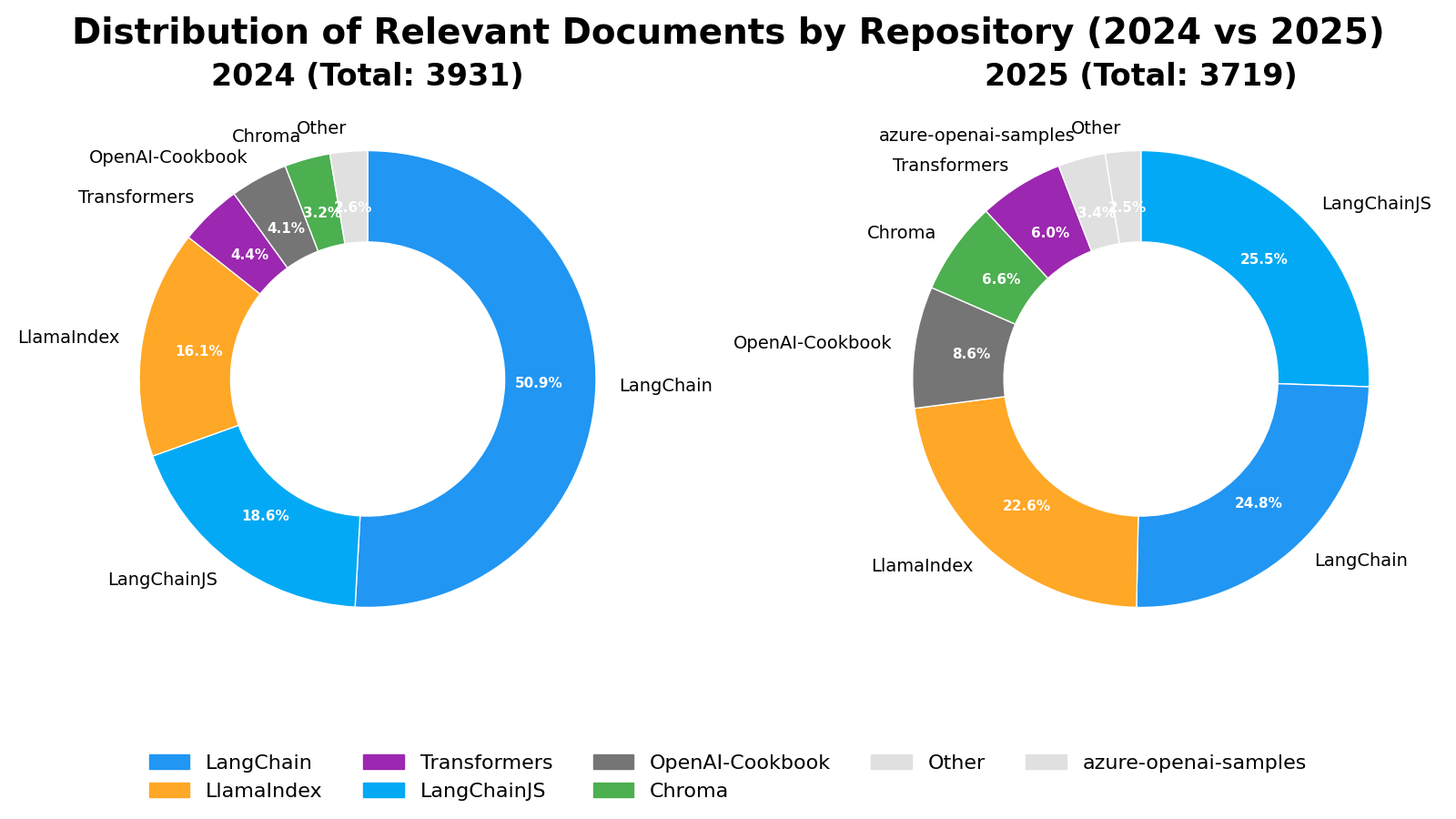}
    \vspace{-0.3cm}
    \caption{An illustration of the distribution of relevant documents (in \%) by each GitHub repository for 2024 and 2025.}
    \label{fig:relevance_distribution_pie}
\end{figure}

\subsection{Nugget Generation}

Nuggets are atomic key facts that are essential for constructing an answer~\cite{lin:2005, lin:2006, pradeep2025greatnuggetrecallautomating, thakur2025freshstack}. 
We automatically generate nuggets from Stack Overflow queries and answers to break down long-form answers and facilitate document-level assessment. 
The nuggets also crucially serve as queries in the oracle retrieval phase~\cite{thakur2025freshstack}. The use of nuggets as a tool for evaluation has grown in the field of retrieval system performance~\cite{10.1145/3726302.3730316, pradeep2025greatnuggetrecallautomating}.  

Similarly, LLMs demonstrate a strong ability to extract relevant and concise information from documents, rivaling the human annotation process, which is cumbersome and time-consuming~\cite{pradeep2025greatnuggetrecallautomating, metropolitansky2025effectiveextractionevaluationfactual}. 
Therefore, for each query in LangChain, we reutilize the nuggets generated using GPT-4o in FreshStack~\cite{thakur2025freshstack}. 

\subsection{Oracle Retrieval}\label{sec:oracle-retrieval}

To create a test collection we require judgment pools. In other words, we must retrieve and annotate documents that are relevant for each query. 
Previous studies \cite{miracl, papadimitriou2024rag, thakur2025freshstack} show the importance of using diverse retrieval models when constructing judgment pools.  

\smallskip
\noindent\textbf{Retrieval Techniques.} First, we utilize BM25~\cite{bm25}, which retrieves lexically similar documents. 
Our next model is BGE (Gemma-2)~\cite{chen:2024}, a dense retriever model\footnote{BGE (Gemma-2): \url{https://huggingface.co/BAAI/bge-multilingual-gemma2}} with an embedding size of 3584 and an 8K context length. 
Next, E5 Mistral (7B)~\cite{wang:2024b} is another dense retriever model\footnote{E5 Mistral 7B: \url{https://huggingface.co/intfloat/e5-mistral-7b-instruct}} with an embedding size of 4096. 
Furthermore, we employ Qwen3 (4B)\footnote{Qwen3 (4B) Embedding: \url{https://huggingface.co/Qwen/Qwen3-Embedding-4B}} which has an embedding size of 4096 and a 32K context length. Qwen3 embeddings achieve state-of-the-art accuracy in retrieval on various domains~\cite{qwen_3:2025}. 
Finally, we normalize the retrieval scores for each model between $[0,1]$ and sum them up in hybrid fusion.

\smallskip
\noindent\textbf{Retrieval Settings.} To further increase diversity in our judgment pools, we use different approaches to construct queries in the oracle setting. 
We begin by using the original answers provided on Stack Overflow. 
Next we use the generated nuggets by GPT-4o, which contains concise and structed information. 
We then generate sub questions using the Qwen3-4B-Instruct model.\footnote{Qwen3-4B-Instruct: \url{https://huggingface.co/Qwen/Qwen3-4B-Instruct-2507}} 
Instead of using the original verbose user query, we use Qwen3 to decompose it into sub-questions, enabling effective retrieval through chain of thought prompting~\cite{10.5555/3600270.3602070}. 
Furthermore, we use Qwen3 to generate the closed-book answer to the Stack Overflow query. 
Finally, we collect the top 50 documents using fusion for each technique described above, and pass them along to the next stage.

\begin{figure}[t]
  \centering
  \includegraphics[width=0.95\columnwidth]{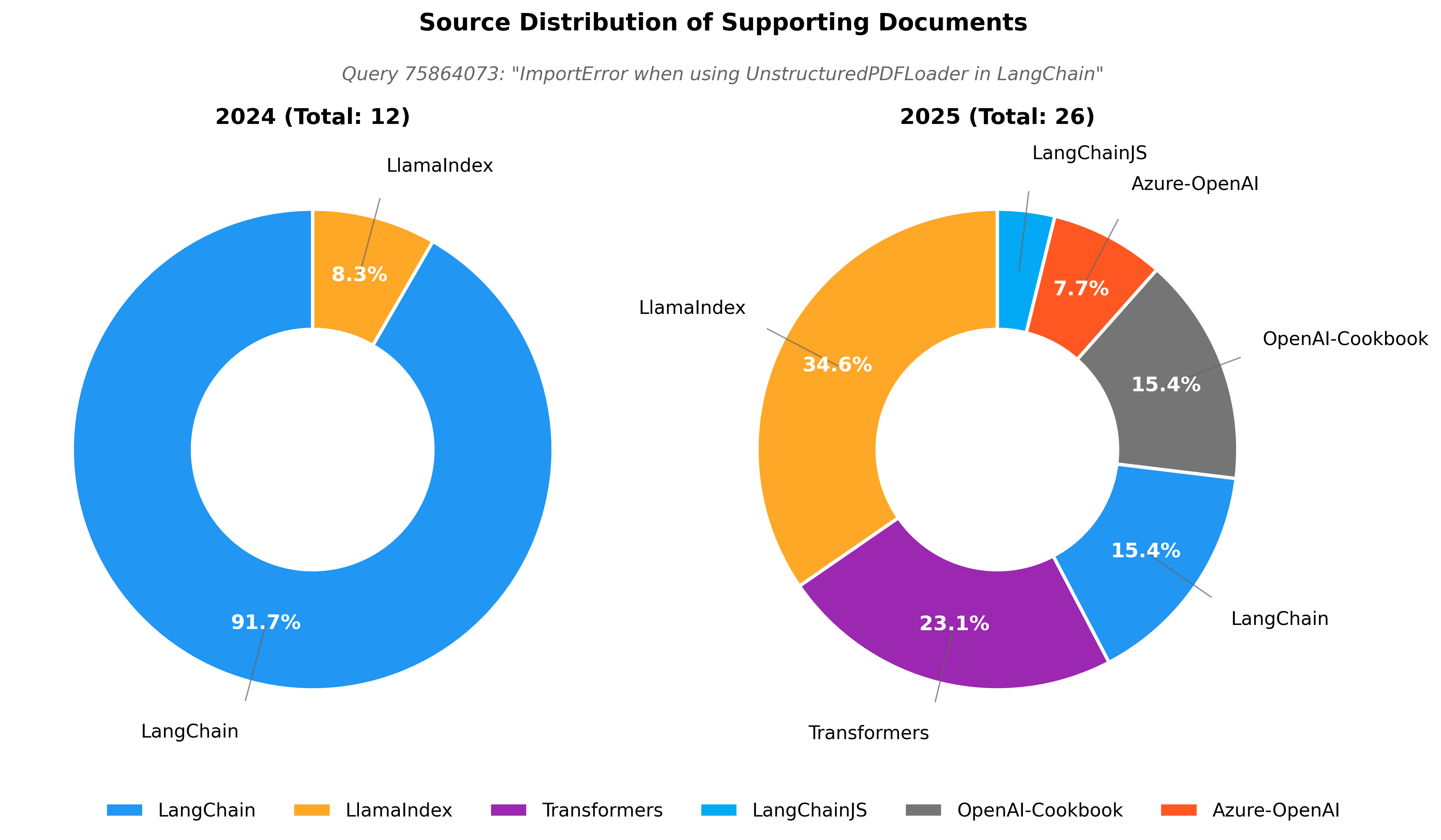}
  \caption{Source distribution shift for LangChain query \texttt{75864073} between 2024 and 2025 corpora snapshots.}
  \label{fig:query_temporal_shift}
\end{figure}

\subsection{Nugget-Level Assessment}

Deciding whether a document is relevant to a given query has traditionally relied on human judges. 
However, this process can be expensive in terms of both cost and time. 
Recent studies have shown potential in utilizing LLMs as automatic judges for relevance judgments~\cite{10.1145/3626772.3657707, rahmani2024llmjudge, 10.1145/3731120.3744591, upadhyay:2024b}. 
IR test collections typically contain short queries, making the judgment task easier for LLMs. Due to verbose and long queries in the Stack Overflow questions, we opt to judge relevance based on documents supporting nuggets instead. 

We utilize \textbf{Cohere's Command A}\footnote{Cohere Command A: \url{https://cohere.com/blog/command-a}} as the judge, with 111B parameters and a 256k context length and has shown performance rivaling GPT-4o~\cite{cohere2025command}. 
We provide Command A with the query and answer (for context), generated nuggets and a set of retrieved documents. Finally, we task it with identifying which documents support information contained within the nuggets. 
In other words, we define a document to be relevant if and only if it supports information for at least one nugget for a given query. 

\section{Experimental Results \& Analysis}

The main focus of our analysis is to compare our corpus and the performance of various models across both snapshots of the selected repositories. 
First, we explore whether LangChain questions asked from 2023 to 2024 can still be answered using the latest 2025 corpus (Section \ref{section:temporal}). 
Subsequently, we expand on some of the observed changes in the GitHub repositories in terms of relevant documents (Section \ref{sec:corpus-temporal}). Analyzing these changes allows us to better understand the complexity of the retrieval task on technical documentation. Finally, we evaluate how diverse retrieval systems perform across both temporal snapshots (Section \ref{sec:benchmark-analysis}). 

\subsection{Temporal Support of Queries}\label{section:temporal}

In this section, we explore \textbf{RQ1}, whether existing queries continue to be grounded in a document corpus with temporal shifts.
For example, ``agents'' used to be part of the LangChain repository in 2024. However, since 2025, the implementation of agentic systems has been moved to langGraph~\cite{Wang2024IntelligentSA}. 
Therefore, queries made on subject of agentic systems \emph{would fail} to retrieve relevant documents from the 2025 LangChain GitHub repository. 

\smallskip
\noindent\textbf{Experimental Setup.} To determine whether dynamic changes in the LangChain corpus limit the set of queries asked previously in time, we conduct the following experiment:
If all nuggets for a query are supported by at least a single relevant document in the corpus, there is sufficient information in the corpus to ground the key facts necessary to construct the Stack Overflow answer.

\smallskip
\noindent\textbf{Results.} 
We run the experiment on both snapshots of our corpus, and observe that only 1 nugget is not supported in the 2025 corpus. 
That is, for all but one of the 640 nuggets (across 203 queries), there exists at least one relevant document in the 2025 corpus to support the information it provides. 
We can thus conclude that both the 2025 and 2024 corpus snapshots can ground existing queries even under temporal shifts, and provide relevant documents required to answer questions in LangChain. 

\begin{figure}
        \centering
        \includegraphics[trim={0 20 10 20}, clip, width=0.48\textwidth]{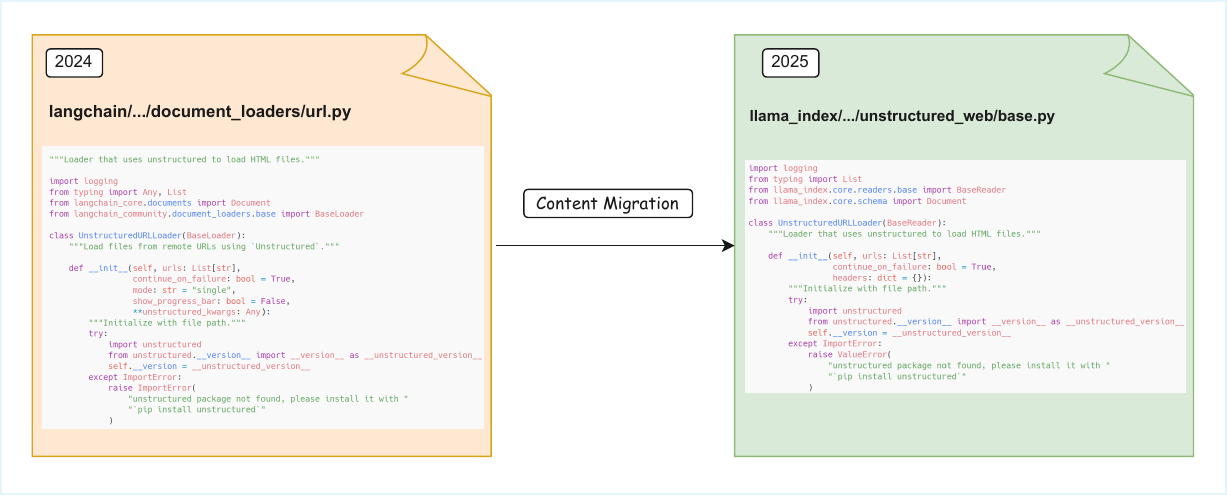}
        \caption{\texttt{UnstructuredURLLoader} class migrated for LangChain query \texttt{75864073} from LangChain (2024) and integrated into LlamaIndex (2025).}
        \label{fig:code_migration_example}
    \end{figure}

\subsection{Corpus Temporal Changes}\label{sec:corpus-temporal}

In this section, we explore \textbf{RQ2}, by analyzing the differences in relevant documents across both snapshots. Previously, we outlined a potential scenario where implementation details (e.g., agents) migrate from one repository to another. 
Therefore, it is important to consider the impact of such shifts occuring within the selected set of repositories in the corpus. 

\smallskip
\noindent\textbf{Experimental Setup.} We examine the following question: among the documents assessed relevant by Command A, which GitHub repositories are they associated with? We show the source distribution of the supporting documents in \autoref{tab:langchain-2024-2025} and \autoref{fig:relevance_distribution_pie}.

\smallskip
\noindent\textbf{Results.} \autoref{fig:relevance_distribution_pie} highlights the underlying distribution shift in document relevance across both snapshots. 
First, we observe a reduction in relevant documents originating from the LangChain repository. 
In 2024, LangChain was the source of 50.9\% of all relevant documents. However, in 2025, this drops down to 24.8\% (just below LangChainJS with 25.5\%).
Furthermore, In 2025, we observe greater diversity in the sources of relevant documents. This shift is consistent with prior work~\cite{Malviya2024Scalability}, which characterizes LangChain as a modular ecosystem that relies on related repositories such as LlamaIndex. 

\begin{table}[t!]
\centering
\caption{Document retrieval on LangChain using the 2024 and 2025 temporal snapshots. Metrics used for evaluation are $\alpha$-nDCG@10, Coverage@20, and Recall@50.}
\label{tab:langchain-2024-2025}
\resizebox{0.47\textwidth}{!}{
\begin{tabular}{l
  ccc | ccc}
\toprule
\multicolumn{1}{l}{\multirow{2}{*}{\textbf{Embedding Model}}} & \multicolumn{3}{c|}{\textbf{LangChain (2024)}} & \multicolumn{3}{c}{\textbf{LangChain (2025)}} \\

\cmidrule(lr{0em}{1em}){2-4}
\cmidrule(lr{0em}{1em}){5-7}

& 
$\alpha$N@10 & C@20 & R@50 &
$\alpha$N@10 & C@20 & R@50 \\
\midrule

BM25~\cite{bm25} & 0.228 & 0.576 & 0.170 & 0.187 & 0.489 & 0.154 \\ \midrule

Qwen3 (8B)~\cite{qwen_3:2025}  & \textbf{0.503} & \textbf{0.877} & 0.432 & 0.480 & 0.868 & 0.424 \\
Qwen3 (4B)~\cite{qwen_3:2025}  & 0.501 & 0.868 & \textbf{0.436} & \textbf{0.482} & \textbf{0.873} & \textbf{0.437} \\
Qwen3 (0.6B)~\cite{qwen_3:2025} & 0.442 & 0.821 & 0.362 & 0.411 & 0.782 & 0.372 \\
\midrule

Stella (1.5B)~\cite{stella:2024} & 0.444 & 0.824 & 0.374 & 0.430 & 0.830 & 0.371 \\
Stella (400M)~\cite{stella:2024} & 0.386 & 0.812 & 0.340 & 0.388 & 0.805 & 0.345 \\
\midrule

Granite-Emb.-R2~\cite{granite:2025} & 0.396 & 0.767 & 0.318 & 0.359 & 0.766 & 0.325 \\
Granite-Emb.-R2 (Small)~\cite{granite:2025} & 0.393 & 0.768 & 0.311 & 0.375 & 0.752 & 0.311 \\
\midrule

GTE (ModernBERT-base)~\cite{li:2023} & 0.365 & 0.716 & 0.271 & 0.343 & 0.728 & 0.279 \\
GTE (large) v1.5~\cite{li:2023}   & 0.324 & 0.709 & 0.239 & 0.276 & 0.683 & 0.242 \\
\midrule

E5 Mistral (7B)~\cite{wang:2024b}  & 0.464 & 0.851 & 0.395 & 0.434 & 0.828 & 0.388 \\
BGE (Gemma-2)~\cite{chen:2024} & 0.423 & 0.794 & 0.341 & 0.401 & 0.828 & 0.348 \\ 
Jina v4~\cite{jina:2025} & 0.408 & 0.811 & 0.381 & 0.417 & 0.821 & 0.379 \\
Voyage-4-nano~\cite{voyage_4} & 0.441 & 0.811 & 0.399 & 0.426 & 0.824 & 0.396 \\
\bottomrule
\end{tabular}}
\end{table}

\smallskip
\noindent\textbf{Case Study.} 
As a case-study, we examine query \texttt{75864073} (title: \emph{ImportError with UnstructuredPDFLoader in LangChain}), shown in \autoref{fig:code_migration_example}. 
First, we observe that the relevant documents increases from 12 to 26, demonstrating that related frameworks increasingly overlap in functionality, resulting in redundancy rather than deferring to the original source. 
In 2024, LangChain is the primary source for relevant documents (91.7\%). However, in 2025, the relevant documents are spread across six repositories, with LlamaIndex (34.6\%) being the largest source. 
As illustrated in \autoref{fig:code_migration_example}, the \texttt{UnstructuredURLLoader} class, a relevant document for query 75864073, was migrated from LangChain to LlamaIndex.

We suspect two reasons for this observation: (1) LangChain reorganized its documentation and removed content about third-party integrations in 2025, and (2) competing frameworks like LlamaIndex and Transformers expanded their documentation to cover similar processing workflows in 2025. 

These observations reiterate the inherent complexity of the retrieval task on technical documentation.
The retrieval system cannot simply find the exact same file across both corpus snapshots to retrieve relevant information. 
Due to temporal changes in the repository, a robust retrieval system must focus on the underlying document structure and content presented at runtime, despite the lack of temporal consistency.

\begin{table}[t!]
\centering
\caption{Kendall $\tau$ correlation for each metric between model rankings in LangChain 2024 and 2025 corpus snapshots.}
\resizebox{0.48\textwidth}{!}{
\begin{tabular}{lccc}
\toprule
\textbf{Metric} &  \textbf{$\alpha$-nDCG@10} & \textbf{Coverage@20} & \textbf{Recall@50} \\ \midrule
Kendall $\tau$ & 0.846 & 0.692 & 0.978 \\
\bottomrule
\end{tabular}}
\label{tab:kendall-tau}
\end{table}

\subsection{Benchmark Analysis}\label{sec:benchmark-analysis}
In this section, we attempt to answer \textbf{RQ3}. We benchmark the retrieval accuracy of diverse embedding models across both snapshots and compare them to observe any changes in model ranking.

\smallskip
\noindent\textbf{Experimental Setup \& Results.} 
In addition to the retrieval models used in Section \ref{sec:oracle-retrieval}, we benchmark models from the following series: Qwen3 embeddings~\cite{qwen_3:2025}, Stella~\cite{stella:2024}, Granite-R2~\cite{granite:2025}, GTE~\cite{gte:2023}, Jina V4~\cite{jina:2025} and Voyage-4-nano~\cite{voyage_4}. 
For evaluation, we follow the setup in FreshStack~\cite{thakur2025freshstack} and use $\alpha$-nDCG@10, Coverage@20, and Recall@50 to measure retrieval diversity and relevance. 
Individual metric definitions are provided in Thakur et al.~\cite{thakur2025freshstack}. The document retrieval results are presented in \autoref{tab:langchain-2024-2025}. 
Overall, we observe that Qwen3 (4B) \& (8B) achieve the highest scores. For $\alpha$-nDCG@10 and Recall@50, we generally observe a decrease in scores between the 2024 and 2025 snapshots for the majority of retrieval models.

\smallskip
\noindent\textbf{Model Ranking Correlation.} 
We utilize the Kendall $\tau$ correlation metric to measure changes in model rankings between the 2024 and 2025 corpus snapshots.
From \autoref{tab:kendall-tau}, we observe a very strong correlation between model rankings across both snapshots, particularly for Recall@50 (Kendall $\tau$ = 0.978) and $\alpha$-nDCG@10 (Kendall $\tau$ = 0.846). 
However, the correlation is relatively weaker for Coverage@20 (Kendall $\tau$ = 0.692).

The positive correlation indicate that models across different temporal snapshots remin fairly consistent in their ability to retrieve \textbf{relevant information}, despite dynamic changes in the corpus. However, they appear to struggle when retrieving diverse relevant passages needed to cover all nuggets required to synthesize a comprehensive Stack Overflow answer. 

\section{Conclusion \& Future Work}

In this work, we investigate how temporal corpus drift affects FreshStack on the LangChain subset. We first observe that ``all except one'' (202/203) queries continue to have supported documents (i.e., relevant documents) for each nugget in the October 2025 corpus snapshot. 
Next, we investigate the reason for the high utility of such queries. 
We found that functions and classes often migrate to other competitor frameworks, e.g., LangChain $\rightarrow$ LlamaIndex. 
Finally, we observe that model rankings across both temporal snapshots of the corpus are strongly correlated, indicating that retrieval systems are consistent in retrieving relevant documents despite changes in the corpus over time. 
These findings suggest that FreshStack (LangChain) is robust to dynamic changes in technical code documentation.

A pivotal aspect of our work is the focus on a test collection built from multiple GitHub repositories as the corpus. 
Due to the way LangChain is structured, much of the code functionality, even when deprecated within one repository, is preserved or migrated to related or competitor codebases. 
However, it is pertinent to consider to what extent other domains will behave similarly. 
For example, in Wikipedia, information within the corpus may change over time, and even the answer to a question may evolve. 
In such contexts, it may be necessary to generate nuggets at each temporal snapshot before constructing the test collection. 
Furthermore, it would be interesting to investigate how consistent the performance of retrieval systems remains across different snapshots in time.

\section*{Acknowledgments}
We would like to thank Cohere for providing API credits to run Command A as the judge in our experiments. This research was supported in part by the Natural Sciences and Engineering Research Council (NSERC) of Canada.

\balance
\printbibliography

\end{document}